\documentclass[12pt]{article}
\usepackage{epsfig}
\begin{document}

\title{\bf Experimental study  of the decay $\phi\to\eta\gamma$
in multi-photon final state}
\author{
M.N.Achasov, S.E.Baru, A.V.Berdyugin, A.V.Bozhenok, A.D.Bukin,\\
D.A.Bukin, S.V.Burdin, T.V.Dimova, S.I.Dolinsky, V.P.Druzhinin,\\
M.S.Dubrovin, I.A.Gaponenko, V.B.Golubev, 
V.N.Ivanchenko\thanks{email:V.N.Ivanchenko@inp.nsk.su}, \\
A.A.Korol, S.V.Koshuba, I.N.Nesterenko, E.V.Pakhtusova, A.A.Polunin, \\
E.E.Pyata, A.A.Salnikov, S.I.Serednyakov, V.V.Shary, Yu.M.Shatunov, \\
V.A.Sidorov, Z.K.Silagadze, Yu.S.Velikzhanin\\
 \\
Budker Institute of Nuclear Physics, Novosibirsk, 630090, Russia
 \\
}
\maketitle

\begin{abstract}
   In the SND experiment at VEPP-2M $e^+e^-$ collider  the
$\phi(1020)\to\eta\gamma\to3\pi^0\gamma$ decay was studied. 
The branching ratio
$B(\phi \to \eta\gamma )=(1.246 \pm 0.025 \pm 0.057)~\%$ 
was measured. 
\end{abstract}


\vspace{3 pt}

The $\phi\to\eta\gamma$ decay is a classical magnetic 
dipole transition from $\phi$ to $\eta$ meson 
which was studied in many experiments \cite{PDG}. In this 
work we describe  the measurement of the $\phi\to\eta\gamma$
branching ratio in the SND experiment at the VEPP-2M 
$e^+e^-$ collider in Novosibirsk.
Spherical non-magnetic detector SND \cite{SND} 
was designed for the experimental
study of $e^+e^-$-annihilation at center of mass energy about $1~GeV$.
Its main part is a three layer electromagnetic calorimeter 
consisted of 1630 NaI(Tl) crystals \cite{CAL,NAI}.
The experiment was performed at VEPP-2M in 1996 \cite{FI96}.
It consists of 6 successive runs 
at 14 different beam
energies in the region $2E_0=(980-1050)~MeV$ covering
the peak and close vicinity of the $\phi$ resonance.
The total integrated luminosity
is equal to $3.7~pb^{-1}$ which corresponds to  $7.6\cdot10^6$ 
$\phi$-mesons produced. The luminosity was determined using 
events of the two-quantum annihilation process.

\begin{figure}[b!] 
\centerline{\epsfig{file=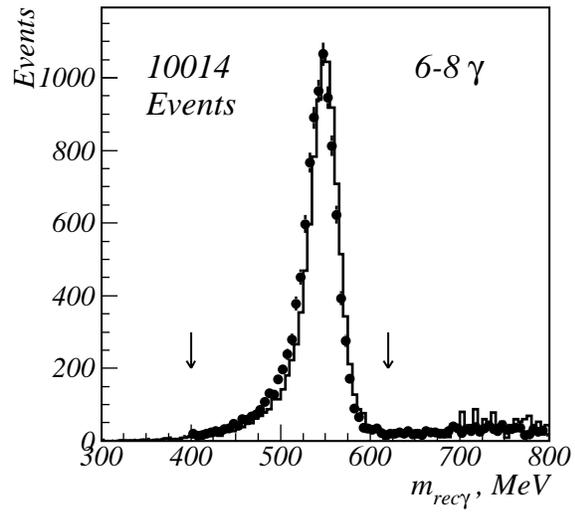}}
\caption{The recoil mass of the most energetic photon in an event.
Points -- data, histogram -- simulation.}
\label{etg98}
\end{figure}

The following reaction was studied:
$$e^+e^-\to\phi\to\eta\gamma\to 3\pi^0\gamma\to7 \gamma.$$ 
The events were selected with 6--8 photons 
emitted at angles of more than 27 
degrees with respect to the beam and no charged particles.
Standard SND cuts \cite{FI96}
 on energy-momentum balance  in an event were used.
 As a result of such selection 
cosmic background is excluded and the 
main background process
$e^+e^-\to\phi\to K_S K_L \to\pi^0\pi^0 K_L$
is suppressed. It is seen from  Fig.\ref{etg98}, where 
spectrum is shown for the recoil 
mass $m_{rec\gamma}$
of the most energetic photon in an event.
The peak on the mass of the $\eta$ meson dominates.
For final selection of $\eta\gamma$ events  a soft
cut $400~MeV < m_{rec\gamma} < 620~MeV$  was used.
The  $\phi\to K_S K_L$ background contribution is about $1~\%$, which
 was evaluated using number of events from the
$m_{rec\gamma}$ interval from 620 to
840~MeV.  It worth noting that 
number and spectrum of  $K_S K_L$ events are
in a good agreement with Monte Carlo simulation.

The number of thus selected events in each energy point for each run
allows to determine  the
visible cross section $\sigma_{vis}$, which was fitted 
for each of 6 runs
separately according to following expression:
$$ \sigma_{vis}(s)=\sigma_0\beta(s)\varepsilon\sqrt{
\frac{1-m_{\eta}^2/s}{1-m_{\eta}^2/m_{\phi}^2}}\cdot
{\left\vert\frac{\Gamma_{\phi}m_{\phi}}
{m_{\phi}^2-s-im_{\phi}\Gamma_{\phi}(s)}+A\right\vert}^2, s=4E^2,$$
where E is a beam energy,
  $\sigma_0$ -- cross section
in the $\phi$-meson pole (free parameter of the fit),
$\beta(s)$ -- radiative correction which was calculated according
to Ref.\cite{KUR},
$\varepsilon$ -- detection efficiency, 
$A$ -- interference term.
In the fit procedure the beam energy spread of $300~keV$ and 
the instability of average beam energy of $100~keV$ were taken into account.
The shifts in the beam energy scale for each run were determined using
the decay mode  $\phi\to K_S K_L$ \cite{FI96}.

The results of the fit are shown in  Table~1.
The detection efficiency is individual for each run 
because of small differences in trigger settings and in  number 
of broken or noisy calorimeter channels.  
The value $A$ describes the interference between $\phi$ meson and 
tails of $\rho$, $\omega$ resonances and with a possible anomaly 
contribution \cite{BEN}.
The results shown in Table~1 were obtained for $A = 0$. If
the fit is performed with $A$ as a free parameter the result coincides 
with the previous assumption.
If $A$ is calculated    
according to Ref.\cite{BEN}, then the results are practically the same.
If standard vector dominance model is used as in the analysis of
previous experiments \cite{DOL}, then value of $\sigma_0$ decreases 
by only $1.5~\%$. We include described model uncertainty in the 
systematic error.

Averaging the data from the Table~1, one can found 
$$\sigma_0=(16.95\pm 0.34\pm0.59)~nb,$$
where the first error is a statistical one and the second is systematic.
Because the results in Table~1 demonstrate some difference, 
the scale factor for these measurements was
calculated according to PDG recommendations. It was found to be 1.5 and 
was taken into account in the presented statistical error.
The systematic error was estimated to be  $3.5~\%$. 
It is mainly determined by the
the systematic uncertainty in normalization ($3~\%$),
the background subtraction error ($1~\%$),
the error in the detection efficiency estimation ($0.5~\%$),
the error in the value of the
interference term  ($1.5~\%$).

Using the following expression:
$$ \sigma_0=\frac{12\pi B(\phi\to\eta\gamma)B(\phi\to e^+e^-)B(\eta\to 3\pi^0)}
{m_{\phi}^2}$$
and PDG data \cite{PDG}
one can find the value of the branching ratio
$$B(\phi\to\eta\gamma)=(1.246\pm 0.025\pm0.057)~\%,$$
where first error is a statistical one and the second is systematic.  Taken
into account were systematics in $\sigma_0$ and in PDG data for
$B(\phi\to e^+e^-)$ ($2.7~\%$) and 
$B(\eta\to3\pi^0)$ ($1.2~\%$).

The result is in agreement with average
PDG value $(1.26\pm0.06)~\%$ and with each previous
experiment \cite{PDG}. 
At the moment it is the most accurate single
measurement of $B(\phi\to\eta\gamma)$ based on $10^4$ selected events.
Precision can be improved if systematic uncertainty is decreased.
Note that the limiting factors for 
this measurement are the accuracy of normalization and the accuracy of the
$\phi\to e^+e^-$ branching ratio. 

\vspace{3mm}
The work is partially supported by RFBR 
(Grants No 96-02-19192, 96-15-96327) and
STP ``Integration'' (Grant No 274).

\begin{table}
\caption{The results of the study of $\phi\to\eta\gamma$
decay for 6 independent runs.}

\vspace {2pt}

\label{IVN:T1}
\begin{tabular}{|l|l|l|l|l|l|}
\hline
Experiment & Events & $N_{\phi}\cdot 10^{-6}$ & $\varepsilon,~\%$ &
$\sigma_0, nb$ & $\chi^2/n_D$ \\
\hline
PHI9601 & 1064 & 0.867 & 32.15 & $16.14\pm0.59$ & 11/8  \\
PHI9602 & 1465 & 1.143 & 32.12 & $16.79\pm0.53$ & 13/8 \\
PHI9603 & 2171 & 1.721 & 31.69 & $16.85\pm0.45$ & 16/10 \\
PHI9604 & 1258 & 1.025 & 31.33 & $16.05\pm0.57$ & 16/11 \\
PHI9605 & 2286 & 1.607 & 32.05 & $17.92\pm0.56$ & 16/11 \\
PHI9606 & 1770 & 1.227 & 31.73 & $18.30\pm0.66$ & 5/9   \\
\hline
\end{tabular}
\end{table}

\end{document}